\documentstyle[amssymb,aps]{revtex}
\begin{document}
\author{A.B. Kobakhidze}
\address{The E.Andronikashvili Institute of Physics, 380077 Tbilisi, Georgia}
\title{The top quark mass in the minimal top condensation model with extra
dimensions}
\draft
\maketitle

\begin{abstract}
The minimal dynamical electroweak symmetry breaking through top condensation
in the presence of compact large extra dimensions is studed. It is shown
that thanks to the power-low evolution of gauge and Yukawa couplings the
original BHL predictions for the top quark mass are significantly lowered
and even for small cut--off scale $\Lambda \sim few$ TeV one can obtain
experimentally allowed values.
\end{abstract}

\pacs{12.60.Rc, 12.25.Mj, 11.10.Hi}

Despite the success of the Standard Model (SM) in describing the
experimental data with impressive accuracy, the physical mechanism behind
the electroweak symmetry breaking (EWSB) and somewhat related explanation of
the masses and mass hierarchies of elementary particles remains as the most
outstanding problem. The top condensation is an interesting mechanism for
the dynamical EWSB (see [1] for review and references therein). The beauty
and strength of this mechanism lies in the fact that usually the set
assumptions are rather limited and that it can explain simultaneously the
dynamical generation of the heavy top quark mass as well as all or part of
the EWSB. A detailed investigation of the minimal scenario, where the EWSB
follows from the top quark condensation alone, was done by Bardeen, Hill and
Lindner (BHL) [2]. Eliminating the SM Higgs sector in favor of the following
local, attractive, Nambu--Jona-Lasinio--type interaction 
\begin{equation}
{\cal L}_{NJL}^{4-\dim .}=G(\overline{Q}_{L}t_{R})(\overline{t}_{R}Q_{L})
\end{equation}
(here $Q_{L}^{T}=(t_{L},b_{L})$ is the third generation left--handed quark
doublet and $t_{R}$ the corresponding right--handed singlet and color and
weak isospin indeces are suppressed), in the fermion buble approximation for
sufficiently attractive $G$ ($G>\frac{8\pi ^{2}}{3}\frac{1}{\Lambda ^{2}}$),
they have obtained a non-trivial solution to the gap equation which relates
the top quark mass $m_{top}$ with an ultrviolet cut-off $\Lambda $ and
four-fermion coupling $G$ 
\begin{equation}
m_{top}^{2}=\frac{\Lambda ^{2}-\frac{8\pi ^{2}}{3}G^{-1}}{\ln (\Lambda
^{2}/m_{top}^{2})}
\end{equation}
and a W-boson mass 
\begin{equation}
m_{W}^{2}=\frac{3g_{2}^{2}}{32\pi ^{2}}m_{top}^{2}\left( \ln (\Lambda
^{2}/m_{top}^{2})+\frac{1}{2}\right)
\end{equation}
For a large cut-off, $\Lambda \simeq 10^{15}$ $GeV$, the buble approximation
predicts the value for the top quark mass, $m_{top}\simeq 163$ $GeV$, while $%
m_{top}\simeq 1$ $TeV$ for smaller cut-off, $\Lambda \simeq 10$ $TeV$. This
requires $G$ to be fine tuned in order to cancel the quadratic cut-off
dependence in (2).

An improvment for the top quark mass can be achieved by considering a
low-energy effective field theory which is just the full SM treating the
composite Higgs field as an elementary degree of freadom. The only
difference is the compositness condition that the Higgs field becomes static
at high ($\sim \Lambda $) energies, i.e. $Z_{H}(\Lambda )=0$, thus
recovering from the SM lagrangian together with equation of motion the
structure of the basic NJL lagrangian (1). This compositness condition means
that the top quark mass is governed by an infrared fixed-point solution of
the renormalization group equations (RGEs) [3]. Unfortunately, the full RGE
analyse gives unacceptable predictions for the top quark mass, $%
m_{top}\simeq 220\div 430$ $GeV$ for $\Lambda \simeq 10^{19}\div 10^{4}$ $%
GeV $ [2]. Thus the minimal BHL model as well as its various modifications
(two Higgs and supersymmetric versions), while being in many aspects
phenomenologically viable [4], has a common drawback predicting the value of
the top quark mass too high to fit the experimental data. So, one is led to
consider more complex symmetry breaking scenarios with more condensates and
more parameters [1,4].

In this paper I consider the minimal BHL model assuming the existence of
extra dimensions with relatively large compactification radii wihch tend to
produce lower values for the top quark mass in agreement with experimental
data even for lower cut-off scale $\Lambda $, thus avoiding the fine tuning
required for large $\Lambda $. Extra space-time dimensions appear naturally
in the string theory, and therefore, such an idea is highly motivated from
the fundamental point of view. Recently, the possibility of large extra
dimensions has received considerable attension and their role has been
explored for gauge coupling unification [5,6], for neutrino mass generation
[7], for suppersymmetry breaking [8], to provide an alternative solution to
the gauge hierarchy problem [9]. Cosmological consequences [10], various
phenomenological issues [11] as well as possible collider signatures [12] of
large radii extra dimension have also been investigated.

The way compact extra dimensions is implemented in practiece is to introduce
towers of Kaluza-Klein (KK) exitations assotiated with gauge and matter
fields. The exact details of the spectrum of KK states are, to some extent,
model dependent. Here I closely follow the models described in [5], where
the extra dimentions are compactified on $S/Z_{2}$ orbifold (a circles
subjected to further identification $y_{\alpha }\rightarrow -y_{\alpha }$, $%
\alpha =1,...\delta $; $\delta $ denotes a number of compact dimensions).
The KK exitations in this case can be decomposed into even $\Phi _{+}(x,y)$
and odd $\Phi _{-}(x,y)$ fields 
\begin{eqnarray}
\Phi _{+}(x,y) &=&\sum_{n_{1}=0}^{\infty }...\sum_{n_{\delta }=0}^{\infty
}\Phi ^{(n_{\alpha })}(x)\cos (n_{\alpha }y_{\alpha }/R)  \nonumber \\
\Phi _{+}(x,y) &=&\sum_{n_{1}=1}^{\infty }...\sum_{n_{\delta }=1}^{\infty
}\Phi ^{(n_{\alpha })}(x)\sin (n_{\alpha }y_{\alpha }/R)
\end{eqnarray}
Here $R$ is the radius of compact dimensions (for simplicity I assume that
all extra dimensions have the same radius). Since the appropriate
transformation of the fields under the $Z_{2}$ parity is determind by
interactions, half of the original KK states may be projected out according
to the $Z_{2}$ parity of the fields. If only the odd tower is left, the zero
mode is missing.

Non-supersymmetric theories can be more straightforwardly embedded into
higher dimensions than those of supersymmetric, because KK states no longer
need to form $N=2$ multiplets as it is usually assumed in the supersymmetric
case. Thus, as a minimal scenario I assume that the gauge bosons ($Z_{2}-$
even) have KK exitations, while the chiral SM fermions ($Z_{2}-$even),
living on the orbifold fixed points, have not. At each KK level this
requires the introduction of an additional scalar fields ($Z_{2}-$odd)
transforming as adjoint representations of each SM gauge symmetry group in
order to make corresponding gauge bosons massive [5]. Following to this
framework, I assume that in four dimensions besides the ordinary NJL
interaction effectively appear an infinite number of four-fermion
interactions 
\begin{equation}
{\cal L}_{NJL}^{(4+\delta )-\dim .}=\sum_{n_{1}=0}^{\infty
}...\sum_{n_{\delta }=0}^{\infty }G^{n_{\alpha }}(\overline{Q}_{L}t_{R})( 
\overline{t}_{R}Q_{L})
\end{equation}
with $G^{n_{\alpha }}=\frac{1}{M_{0}^{2}+(n_{\alpha }^{2}/R^{2})}$. The set
of four-fermion interactions in (5) can be viewed as a result of integration
over some heavy ($M_{0}^{2}$) state and its KK exitations ($M_{0}^{2}+\frac{
n_{\alpha }^{2}}{R^{2}}$) [13]. Introducing an auxilary fields $H^{n_{\alpha
}}=$ $G^{n_{\alpha }}(\overline{t}_{R}Q_{L})$ one can rewrite lagrangian (5)
in the equivalent form: 
\begin{equation}
{\cal L}_{NJL}^{(4+\delta )-\dim .}=\sum_{n_{1}=0}^{\infty
}...\sum_{n_{\delta }=0}^{\infty }\left[ -(M_{0}^{2}+\frac{n_{\alpha }^{2}}{
R^{2}})\left| H^{n_{\alpha }}\right| ^{2}+(\overline{Q}_{L}t_{R})H^{n_{
\alpha }}+h.c.\right]
\end{equation}
The set of static fields $H^{n_{\alpha }}$ acquire gauge invariant kinetic
and self-interacting terms after radiative corrections are taken into
account. So below the cut-off scale $\Lambda $ one obtains the SM lagrangian
describing the interactions of gauge and Higgs bosons and their KK
exitations with each other and with chiral fermions living on the orbifold
fixed points: 
\begin{eqnarray}
{\cal L}_{NJL}^{(4+\delta )-\dim .} &=&{\cal L}_{kin.}^{gauge}+{\cal L}
_{kin.}^{fermion}+\sum_{n_{1}=0}^{\infty }...\sum_{n_{\delta }=0}^{\infty
}\allowbreak [Z_{H}\left| D_{\mu }H^{n_{\alpha }}\right|
^{2}-M_{H^{n_{\alpha }}}^{2}\left| H^{n_{\alpha }}\right| ^{2}  \nonumber \\
&&-\frac{1}{2}\lambda (H^{n_{\alpha }+}H^{n_{\alpha }})^{2}+\left( ( 
\overline{Q}_{L}t_{R})H^{n_{\alpha }}+h.c.\right) ]
\end{eqnarray}
whre ${\cal L}_{kin.}^{gauge}$and ${\cal L}_{kin.}^{fermion}$ are the gauge
and fermionic kinetic terms, respectively, and $D_{\mu }=\partial _{\mu
}-\sum_{n_{1}=0}^{\infty }...\sum_{n_{\delta }=0}^{\infty }\left( ig_{2}\tau
^{a}A_{\mu a}^{n_{\alpha }}+ig_{1}B_{\mu }^{n_{\alpha }}\right) $. In a full
analogy with [2] one must demand $Z_{H}\rightarrow 0$ or, equivalently,
rescaling Higgs field and its KK exitations $H^{n_{\alpha }}=\frac{
h^{n_{\alpha }}}{\sqrt{Z_{H}}}$ in order to normalize canonically kinetic
terms in (6), top-Yukawa coupling 
\begin{equation}
g_{t}=\frac{1}{Z_{H}}\rightarrow \infty
\end{equation}
as energy approches to $\Lambda $.

Now let's eximine how the BHL predictions for the top quark mass are changed
in the presence of extra dimensions. First, note that the since the
fermions, and particularly top quark, in our minimal approach have no KK
exitations, it is obvious, that in the fermion buble approximation the
results for the top quark mass are exactly the same as in the BHL model (see
(2) with $G\equiv G^{n_{\delta }=0}$). The RGE improvment, however, is
expected to be drastically different because of contributions of KK states
to $\beta $-- and $\gamma $-- functions, leading to power--low runing of
gauge and Yukawa couplings (for the proper treatment of RGEs in extra
dimensions see [5]) and corresponding changes to the fixed-point solutions
of RGEs (see [14] for the fixed-point solutions in supersymmetric theories
with large extra dimensions).

The one-loop diagrams contributing to the anomalous dimensions of the top
quark consists of top-Higgs and top-gauge internal states. Since in our
minimal approach top quark has no KK exitations each time one pass Higgs or
gauge KK threshold the diagrams contribute as the equivalent SM diagrams.
This is because the KK number is not conserved in the vortices since
translational invariance is broken in the extra dimensions. The Higgs--gauge
loops contributing to the anomalous dimension of the Higgs field also give
the same contribution as an equivalent SM diagrams since now KK number must
be conserved at vortices. The only contribution of KK modes to gauge
coupling $\beta $--functions are came from diagrams with gauge loops and
loops of $Z_{2}$--odd adjoint scalars. Thus, above the energy $\mu
_{0}\approx \frac{1}{R}$ the RGEs for the top-Yukawa $Y_{t}\equiv \frac{
g_{t}^{2}}{4\pi }$ and gauge couplings $\alpha _{i}$ ($i=1,2,3$) are 
\begin{eqnarray}
\frac{dY_{t}}{d\ln (\mu )} &=&\frac{3}{2\pi }Y_{t}^{2}+\frac{Y_{t}}{4\pi }
\left[ \frac{3}{2}Y_{t}-c_{i}\alpha _{i}\right] \frac{d{\cal J}}{d\ln (\mu )}
\nonumber \\
\frac{d\alpha _{i}}{d\ln (\mu )} &=&\frac{b_{i}-b_{i}^{^{\prime }}}{2\pi }
\alpha _{i}^{2}+\frac{b_{i}^{^{\prime }}}{2\pi }\alpha _{i}^{2}\frac{d{\cal %
J }}{d\ln (\mu )}
\end{eqnarray}
where 
\begin{eqnarray}
b_{i} &=&\left( \frac{41}{10},-\frac{19}{6},-7\right)  \nonumber \\
b_{i}^{^{\prime }} &=&\left( \frac{1}{10},-\frac{41}{6},-\frac{21}{2}\right)
\nonumber \\
c_{i} &=&\left( \frac{17}{12},\frac{9}{4},8\right)
\end{eqnarray}
${\cal J}$ in (9) is the integral of eliptic Jacobi theta function [5] 
\begin{eqnarray}
{\cal J}(\mu /\mu _{0},\delta ) &=&\int_{r/(\mu /\mu _{0})^{2}}^{r}\frac{dx}{
x}\left[ \vartheta _{3}(0,e^{-x})\right] ^{\delta },  \nonumber \\
\vartheta _{3}(u,q) &=&\sum_{n=-\infty }^{+\infty }q^{n^{2}}e^{i2nu}
\end{eqnarray}
where $r=\left[ \Gamma (1+\delta /2)\right] ^{\delta /2}$ ($\Gamma $ is the
Euler gamma function). Below the compactification scale $\mu _{0}$
top--Yukawa and gauge couplings evaluate according to the usual
four--dimensional SM RGEs, wich easily recovered from (9) taking $\delta =0$
or, alternatively, $\frac{\mu }{\mu _{0}}=1$ ($\frac{d{\cal J}(\mu /\mu
_{0},0)}{d\ln \mu }=\frac{d{\cal J}(1,\delta )}{d\ln \mu }=2$).

In FIG. 1 I have plotted the dependence of the top quark mass $%
m_{top}=g_{t}v $, $v\simeq 174$ GeV on $\Lambda /\mu _{0}$ solving the full
set of Eqns. (9) numerically with compositness condition (8) for $\Lambda
=10^{4}$, $10^{7}$, $10^{13}$, $10^{19}$ GeV and for $\delta =$1, 2, 4, 7.
The values $m_{top}(0)$ ($\Lambda =\mu _{0}$) in Figure 1 are clearly the
BHL predictions $m_{top}(0)=$ $m_{top}^{BHL}$. When the radius of extra
dimensions is close to $\Lambda $ ($\Lambda /\mu _{0}\lesssim 2.5\div 10^{4}$
for $\delta =7\div 1$) $m_{top}$ is further increased since the gauge
contributions to the evolution of top--Yukawa coupling $Y_{t}$ are more
significant than in the case of four--dimensional SM. This is the direct
consequence of our minimal approache when the chiral fermions are assumed to
live on the orbifold fixed points and thus do not feel the extra dimensions.
However, for larger {\footnotesize \ }$\Lambda /\mu _{0}$ $m_{top}$ is
quikly decreased and one can get the values of top quark mass in the
experimentally allowed range even for $\Lambda =10^{4}$ GeV , $\delta \geq 5$
(demanding $\mu _{0}\geq 1$ TeV). This happens because of power--low ($\frac{
1}{\delta }(\frac{\Lambda }{\mu _{0}})^{\delta }$) running of $Y_{t}$ (in
contrast to the logarithmic running ($\ln (\frac{\Lambda }{\mu _{0}})$) in
the SM) [5, 13] leading to extremely small values of $Y_{t}$ when $\Lambda
/\mu _{0}$ and/or $\delta $ increase even for the small cut--off $\Lambda
\sim $ $few$ TeV. Thus, the problem of quadratic divergenses can be
potentially solved within our framework.

To conclude, I have studed the minimal BHL model for the dynamical
electroweak symmetry breaking assuming the existence of compact extra
dimensions with relatively large radii. It was shown that owing to the
power-low evolution of gauge and Yukawa couplings the original BHL
predictions for the top quark mass are significantly lowered and, even for
small cut--off scale $\Lambda =few$ TeV, one can obtain experimentally
allowed values $m_{top}$, provided the number of extra dimensions with $%
R\approx 1$ TeV$^{-1}$ to be at least 5. This offers an exciting possibility
for future colliders to probe not only the composite nature of the Higgs
boson but also the structure of space--time.

I would like to thank Z.Berezhiani, G.Dvali, J.Chkareuli, I.Gogoladze and
Z.Tavartkiladze for usefull disscussions on various aspects of higher
dimensional theories. This work was partially supported by the INTAS--RFBR
Grant No. 95-567, the INTAS Grant No. 96-155, the grant No. 2-10 of the
Georgian Academy of Sciences and the Georgian Young Scientists Presidential
Award.

\newpage

{\bf Figure caption}

\bigskip

FIG. 1. The top quark mass as a function of $\Lambda /\mu _{0}$, for for $%
\Lambda =10^{4}$, $10^{7}$, $10^{13}$, $10^{19}$ GeV and for $\delta =$1, 2,
4, 7. Intersection of the curves at $\Lambda =\mu _{0}$ corresponds to BHL
predictions, while leftgoing arrow indicates the central experimental value
for the top quark mass.


\begin{references}
\bibitem{}  G. Cveti\v {c}, {\it Top quark condensation -- a review}, eprint
hep-ph/9702381.

\bibitem{}  W. Bardeen, C. Hill and M. Lindner, Phys. Rev. D{\bf 41}, 1647
(1990).

\bibitem{}  B. Pendleton and G. Ross, Phys. Lett. {\bf 98B}, 291 (1981); 
\newline
C. Hill, Phys. Rev. D{\bf 24}, 691 (1981).

\bibitem{}  M. Lindner and E. Schnapka, {\it Dynamical electroweak symmetry
breaking with a Standard Model limit, }in: ``Heavy Flavours II'', eds.
A.J.Buras and M.Lindner, Advanced Series on Directions in High Ehergy
Physics, World Scientific Co., Singapore.

\bibitem{}  K.R. Dienes, E. Dudas and T. Gherghetta, Phys. Lett. B{\bf 436},
(1998) 55; Nucl. Phys. B{\bf 537}, (1999) 47.

\bibitem{}  D. Ghilencea and G.G. Ross, Phys. Lett. B{\bf 442}, 165 (1998); 
\newline
C. Bachas, {\it Unification with low string scale}, eprint hep-ph/9807415; 
\newline
P.H. Frampton and A. Rasin, {\it Unification with Enlarged Kaluza--Klein
Dimensions}, eprint hep-ph/9903479.

\bibitem{}  N. Arkani-Hamed, S. Dimopoulos, G. Dvali and J. March-Russell, 
{\it Neutrino masses from large extra dimensions}, eprint hep-ph/9811448; 
\newline
K.R. Dienes, E. Dudas and T. Gherghetta, {\it Neutrino oscillations without
neutrino masses or heavy mass scales: A higher-dimensional see-saw mechanism}
, eprint hep-ph/9811428.

\bibitem{}  E.A. Mirabelli and M. Peskin, Phys. Rev. D{\bf 58}, 6502 (1998).

\bibitem{}  N. Arkani-Hamed, S. Dimopoulos and G. Dvali, Phys. Lett. B{\bf %
429}, 506 (1998);\newline
H. Hatanaki, T. Inami and C.S. Lim, Mod. Phys. Lett. A{\bf 13}, (1998) 2601; 
\newline
I. Antoniadis and C. Bachas, {\it Branes and the gauge hierarchy}, eprint
hep-th/9812093.

\bibitem{}  K. Benakli and S. Davidson, {\it Baryogenesis in Models with a
Low quantum Gravity Scale}, eprint hep-ph/9810280;\newline
K.R. Dienes, E. Dudas and T. Gherghetta and A. Riotto, {\it Cosmological
Phase Transitions and Radius Stabilization in Higher Dimensions}, eprint
hep-ph/9809406;\newline
G. Dvali and S.--H. H. Tye, {\it Brane Inflation}, eprint hep-ph/9812483.

\bibitem{}  N. Arkani-Hamed, S. Dimopoulos and G. Dvali, {\it Phenomenology,
Astrophysics and Cosmology of Theories with Sub-Millimeter Dimensions and
TeV Scale Quantum Gravity}, eprint hep-ph/9807344;\newline
K. Benakli, {\it Phenomenology of Low Quantum Gravity Scale Models}, eprint
hep-ph/9809582;\newline
Z. Berezhiani and G. Dvali, {\it Flavour Violation in Theories with TeV
Scale Quantum Gravity}, eprint hep-ph/9811378.

\bibitem{}  I. Antoniadis, K. Benakli and M. Quiros, Phys. Lett. B{\bf 331},
313 (1994);\newline
G.F. Giudice, R. Ratazzi and J.D. Wells, Nucl. Phys. B{\bf 544}, 3 (1999); 
\newline
E.A. Mirabelli, M. Perelstein and M. Peskin, {\it Collider Signatures of New
Large Space Dimensions}, eprint hep-ph/9811337;\newline
T. Han, J.D. Lykken and R.--J. Zhang, {\it On Kaluza-Klein States from Large
Extra Dimensions}, eprint hep-ph/9811350;\newline
J.L. Hewett, {\it Indirect Collider Signals for Extra Dimensions}, eprint
hep-ph/9811356.

\bibitem{}  B.A. Dobrescu, {\it Electroweak Symmetry Breaking as a
Consequence of Compact Dimensions, }eprint hep-ph/9812349; {\it Higgs
Compositness from Top Dynamics and Extra Dimensions}, eprint hep-ph/9903407.

\bibitem{}  M. Lanzagorta and G.G. Ross, Phys. Lett. B{\bf 98}, 319 (1995); 
\newline
S.A. Abel and S.F. King, {\it On the fixed points and fermion mass structure
from large extra dimensions}, eprint hep-ph/9809467.
\end{references}
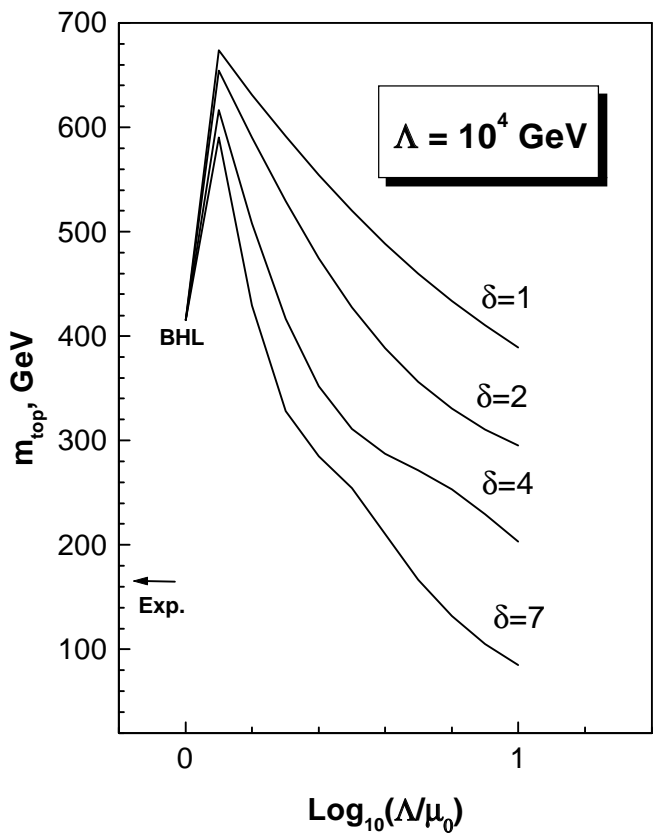
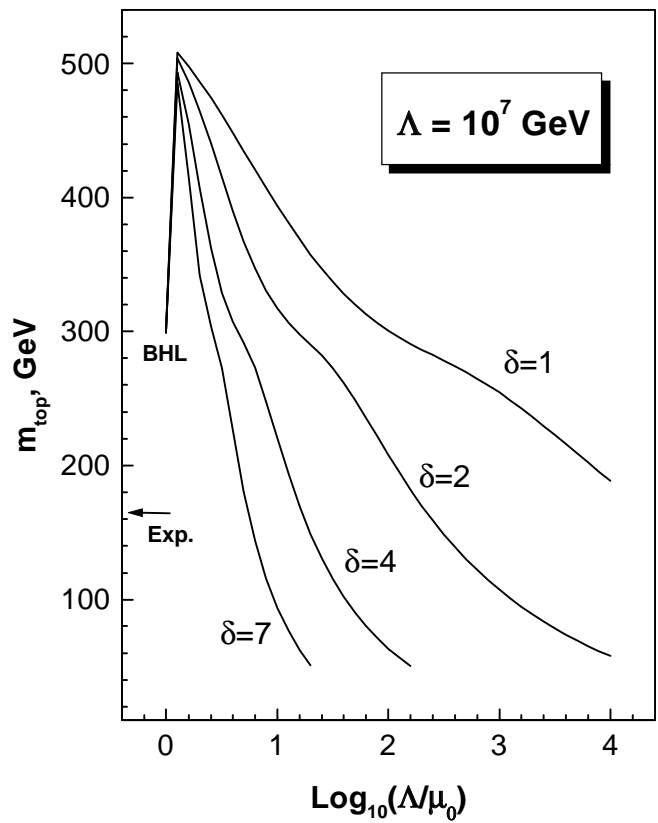
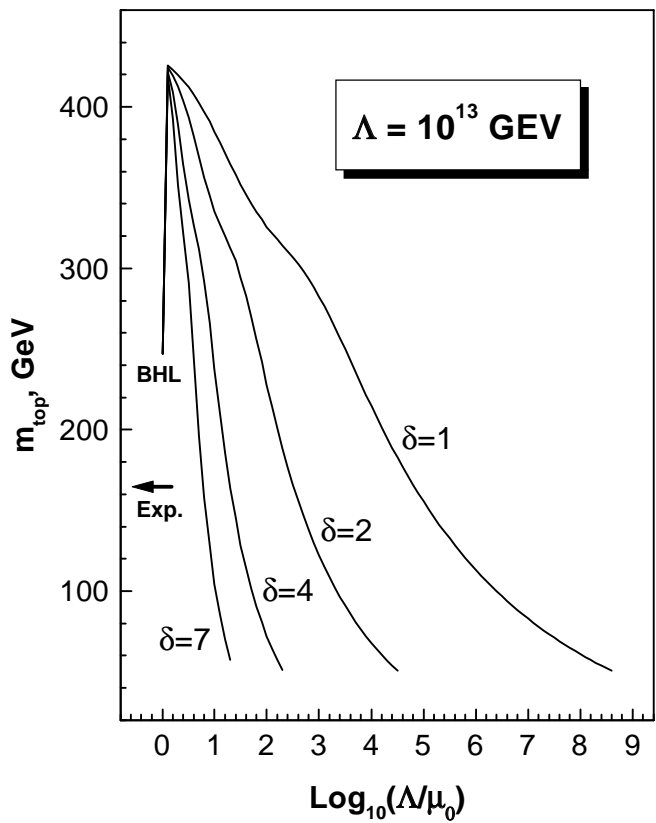
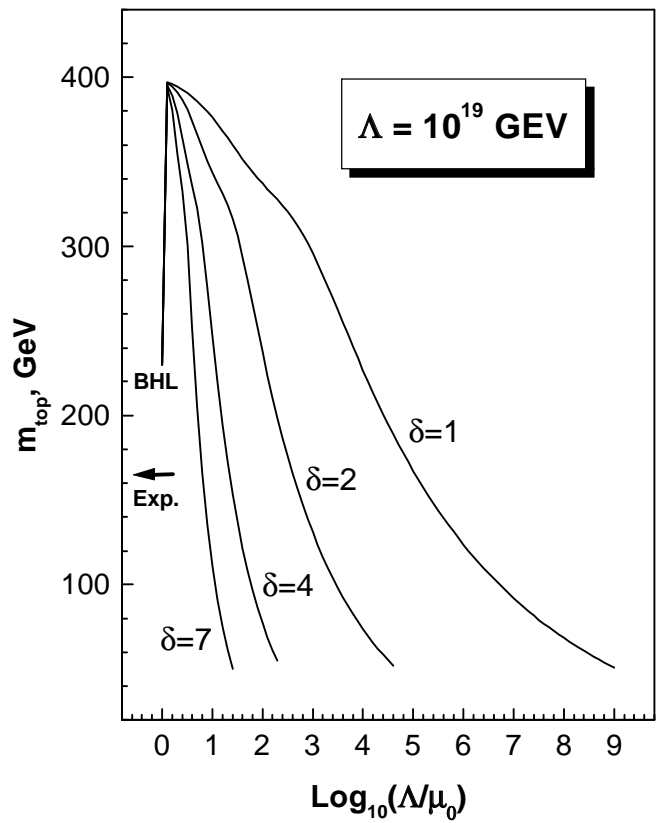
\end{document}